# Improving the Numerical Robustness of Sphere Swept Collision Detection

Jeff Linahan

*Abstract*—This paper discusses improvements to the numerical robustness of the algorithm described in Kasper Fauerby's "*Improved Collision Detection and Response.*" The algorithm addresses a common collision detection query: a moving sphere or ellipsoid vs. a set of motionless triangles. In its current form, the algorithm allows the sphere to penetrate the triangles. The sphere also displays "jittering" behavior when colliding with certain geometry. Most of these problems are the product of insufficient attention to *numerical robustness*, the focus of this paper. Motivated by the importance of numerical robustness in collision detection code, this paper addresses these problems in detail and proposes efficient solutions to them.

## I. INTRODUCTION

SLIDING *collision detection* refers to the process of smoothly sliding a moving entity along any obstacles encountered. This is an attractive feature in any game, but is a difficult technique to implement. Fauerby [2] presents an algorithm that handles sliding collision detection of a sphere against a set of arbitrary meshes stored as polygons. However, insufficient attention was given to the numerical robustness of this algorithm, causing the sphere to sometimes penetrate obstacles. In fact, there seems to be a lack of numerically robust collision detection algorithms in the literature, in spite of their critical importance to computer graphics applications. Fauerby's algorithm is based partly on preceding work with sliding collision detection done by Nettle [4] and Rouwé [5].

We start in section II with an overview of Fauerby's original algorithm. We continue in section III with an outline of an improved algorithm and present the pseudocode of a robust implementation, which eliminates all numerical problems present in Fauerby's implementation [2]. A program is *numerically robust* if it is free from problems in its numerical calculations that cause it to crash, go into infinite loops, or return incorrect results ([1], p. 427). Floating point round-off errors can cause collision detection code to incorrectly allow two objects to intersect, among other things. These types of problems arise from the fact that a digital computer does not have unlimited digits to represent real numbers. These problems can be difficult to debug because our equations may appear mathematically correct, yet their actual coding may contain bugs. To make calculations numerically robust, we must sometimes reorganize our equations, which may potentially result in less straightforward code.

Our presentation is based on the assumption that the reader is familiar with the notion of vectors and operations on vectors. More information about specific topics (e.g., the plane class) can be found in the appendices of Fauerby's paper [2]. Some notational convention: vectors are written in **bold**, whereas points and scalars will be written in *italics*. The normal of a vector **u** is denoted $\hat{\mathbf{u}}$, and the length of **u** is denoted $\|\mathbf{u}\|$. The function *plane_dist*(*p*, *c*) is defined as the distance between the plane *p* and the point *c*. Although the actual algorithm works in 3D space, the figures are drawn in 2D for simplicity and clarity.

## II. BACKGROUND

In this section, we present the background information necessary to understand our work from section III. We briefly discuss the bounding volumes and spatial partitioning techniques commonly used in simplifying the geometry of a graphics scene, then move on to presenting the details of Fauerby's algorithm from [2], upon which our algorithm from section III is built.

### A. Bounding Volumes and Spatial Partitioning

Game characters and other objects in a 3D game are typically represented by a complex graphical polygon mesh enclosed in a much simpler bounding volume, used for collision detection. Among the most popular bounding volumes are spheres, ellipsoids, boxes, cylinders, capsules, and convex polytopes constructed from *k* planes (*k*-DOPs) for some k > 6, because of their cheap intersection tests, small memory footprint, and tight fit ([1], p. 76). Fauerby's algorithm is concerned with a complex intersection test: a moving sphere (or ellipsoid) versus a set of static triangles (also known as a *mesh* or *polygon soup*). Curved bounding volumes are a good choice for game characters and other objects because they provide natural and smooth sliding collision detection. The polygon soup is a static environment, typically composed of a terrain and motionless obstacles (e.g., walls). To simplify the coding of intersection tests, we triangulate polygons that are not triangles.

For motion to look smooth, a game must draw graphics at a rate of 30 to 60 frames per second. This leaves little time for collision detection, let alone all the other calculations required. Most games store the environment

Department of Computer Science, Villanova University, Villanova, PA 19085, USA

jlinah01@villanova.edu

geometry in a spatial partitioning data structure, such as a grid or *k*-dimensional tree as a divide-and-conquer strategy to avoiding testing triangles not in the player's immediate vicinity. These data structures can reduce the time needed to perform collision queries from O(n) to O(log n) or better (where n is the number of triangles) depending on what memory/speed tradeoffs are made ([1], p. 285). Clearly, use of an appropriate partitioning scheme is crucial to a real time implementation of Fauerby's algorithm.

### B. Overview of Original Algorithm

In this section, we present Fauerby's algorithm from [2]. At the beginning of each frame, the sphere (or ellipsoid) starts at a source position and tries to move along a velocity vector to its desired destination, possibly colliding with something along the way. If there are no collisions, the formula for the destination is simple:

$$dest = source + \mathbf{vel} \qquad (1)$$

Here *source* and *dest* represent the center point of the sphere (or ellipsoid) at the start and destination position. This formula can also be expressed parametrically, and we will be referring to values of time $t \in [0, 1]$ throughout the text.

$$dest = source + \mathbf{vel} * t \qquad (2)$$

We need to determine which triangle the sphere or ellipsoid will collide with, if any, along the way. We also need to respond correctly in the case of a collision. One solution would be to stop the sphere from moving upon hitting an obstacle, but to mimic reality closely we prefer sliding collision. Fauerby's algorithm consists of the following steps:

1. Obtain a list of polygons that the sphere or ellipsoid might intersect as it travels from its current position along its velocity vector from the spatial partitioning data structure.
2. Find the transformation matrix that would turn the ellipsoid into a unit sphere and apply it to the ellipsoid and polygons. This step simplifies intersection calculations from ellipsoid vs. triangle to unit sphere vs. triangle.
3. Determine the first triangle, if any, the sphere will intersect if allowed to travel freely. (If no triangle exists, skip to step 8).
4. Move the sphere as far as possible along its velocity vector so it just touches the sliding plane
5. Calculate the sliding plane (sometimes this plane contains the triangle itself).
6. Project the remainder of the velocity vector onto the sliding plane to obtain a new velocity and destination.
7. Goto step 3 and repeat with the new velocity and destination. (We must check again because there is no guarantee we can move along the projected velocity freely.)
8. Move the sphere to its destination. Apply the inverse of the transformation matrix computed in step 2 to the results to undo the trick used to simplify calculations.

Although Fauerby devotes a great deal of his paper to step 3, ([2], p. 9) we will not be discussing it. For our purposes, it suffices to know that this step calculates the time $t \in [0, 1]$ when the sphere first hits a triangle, the *intersection point*, and other statistics on the collision. We will refer to this as the *collision detection step*, which Rouwé [5] explains in detail. There is room for improving the numerical robustness in Fauerby's version of this step, though. For example, a more robust method of computing quadratic roots could be used ([3], p. 411). However, all problems we are concerned with occur in the *collision response step*, which encompasses steps 4 to 7. Fauerby's version of the collision response step is based partly on the ideas presented in Nettle [4], but here we will continue to revise and improve this step.

### C. Overview of Collision Response Step

We start by presenting Fauerby's original code for the collision response step [2]. The main function, *collideWithWorld* (presented in Table I) uses an application-specific function *checkCollision* at line 9 to obtain the list of nearby polygons. This *checkCollision* function performs the calculations of the *collision detection step* on all polygons (triangles) in this list. It then fills the *collisionPackage* struct with statistics about the first collision that will occur (i.e., the collision with the lowest time value *t*). If no collision will occur, *collideWithWorld* simply moves the sphere along the velocity vector (line 11) and returns. Assuming there is a collision, *collideWithWorld* will move the sphere arbitrarily close to the triangle, but only if it is not already very close (lines 14 to 19). Fauerby explains that the sphere should not actually touch the triangle, but should instead be placed very close to it. This is necessary because of floating point rounding errors: if a unit sphere is exactly one unit away from a triangle, a static collision test may report that the sphere is actually intersecting the geometry, touching it at a single point, or separate from it. To solve this problem, Fauerby shortens the velocity vector at line 16.

TABLE I
Collision Response Step from [2]

```
VECTOR CharacterEntity::collideWithWorld(const
VECTOR& pos, VECTOR& vel) const

// All hard-coded distances in this function are
scaled to fit the setting above.
1. float unitScale = unitsPerMeter / 100.0f;
2. float veryCloseDistance = 0.005f * unitScale;

// do we need to worry?
3. if (collisionRecursionDepth > 5) return pos;

// Ok, we need to worry:
4. collisionPackage->velocity = vel;
5. collisionPackage->normalizedVelocity = vel;
6. collisionPackage-
>normalizedVelocity.normalize();
7. collisionPackage->basePoint = pos;
8. collisionPackage->foundCollision = false;

// Check for collision (calls the collision
```

```
routines) Application specific!!
9. world->checkCollision(collisionPackage);

// If no collision we just move along the velocity
10. if (collisionPackage->foundCollision == false)
11.    return pos + vel;

// *** Collision occured ***

// The original destination point
12. VECTOR destinationPoint = pos + vel;
13. VECTOR newBasePoint = pos;

// Update only if we are not already very close,
and if so move very close to intersection … not to
the exact spot.
14. if (collisionPackage-
>nearestDistance>=veryCloseDistance)
{
15.   VECTOR V = vel;
16.   V.SetLength(collisionPackage->nearestDistance-
veryCloseDistance);
17.   newBasePoint = collisionPackage->basePoint +
V;

  // Adjust polygon intersection point (so sliding
plane will be unaffected by the fact that we move
slightly less than collision tells us)
18.   V.normalize();
19.   collisionPackage->intersectionPoint -=
veryCloseDistance * V;
}

// Determine the sliding plane
20. VECTOR slidePlaneOrigin = collisionPackage-
>intersectionPoint;
21. VECTOR slidePlaneNormal =
          newBasePoint-collisionPackage-
>intersectionPoint;
22. slidePlaneNormal.normalize();
23. PLANE
slidingPlane(slidePlaneOrigin,slidePlaneNormal);

24. VECTOR newDestinationPoint = destinationPoint -

slidingPlane.signedDistanceTo(destinationPoint)*sli
dePlaneNormal;

// Generate the slide vector, which will become our
new velocity vector for the next iteration
25. VECTOR newVelocityVector = newDestinationPoint
-
collisionPackage->intersectionPoint;

// Recurse:

// Don't recurse if the new velocity is very small
26. if (newVelocityVector.length() <
veryCloseDistance)
27.    return newBasePoint;

28. collisionRecursionDepth++;
29. return
collideWithWorld(newBasePoint,newVelocityVector);
```

One may wonder why, when the sphere is so close to the triangle that we choose to not move it any closer, we don't simply return from the function early. The intuition here is that even though the sphere is not moving during the current iteration, it is still moving during the current frame, so the velocity vector is still necessary to decide the direction in which the sphere slides during the next iteration of *collideWithWorld*. If we return too early, the sphere will not slide enough, causing the algorithm to degenerate to the unacceptable alternative of stopping the sphere at obstacles, discussed in subsection B.

### D. Finding the Sliding Plane

We now describe how the sliding plane is determined in lines 20 - 23 of the algorithm from Table I. As depicted in Fig. 1, the sliding plane may or may not be parallel to the triangle with which the sphere is colliding.

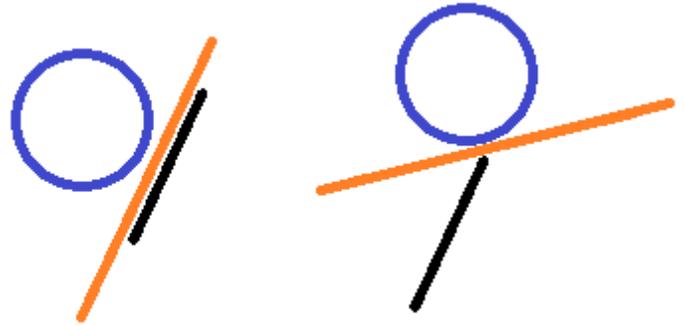

Fig. 1 Sliding plane parallel (left) or not parallel (right) to the triangle. We use 2D representations here: the longer line is the plane, and the shorter line is the triangle.

Because a point and a normal uniquely define a plane, the *intersection point* can be used as a point in the plane, and the vector pointing from the *intersection point* to the center of the unit sphere at time *t* of the collision as the normal. But, because we stopped the sphere at *newBasePoint* (line 17) a moment before it hits the triangle, we must modify the *intersection point*, (line 19) moving it in the direction of **–vel** by *veryCloseDistance* to compensate. This ensures that the two points remain one unit apart and the sliding plane is computed correctly. Fauerby then projects the desired destination onto the sliding plane (line 24). Although Fauerby calls this the *newDestinationPoint*, it actually lies on the sliding plane. If we actually moved the sphere to this point, the plane would bisect it. Fortunately, we are still able the compute the correct projected velocity vector (parallel to the sliding plane) by subtracting *newDestinationPoint* from the modified *intersectionPoint*, which also lies on the plane (line 25). With the values for *newBasePoint* and *newVelocityVector* computed, *collideWithWorld* calls itself recursively (line 29) and the process repeats again.

### III. NUMERICAL ROBUSTNESS PROBLEMS AND SOLUTIONS

In this section, we present ways to simplify and improve the numerical robustness of Fauerby's original collision detection algorithm.

### A. A New Way of Finding the Sliding Plane

Modifying the intersection point succeeds at ensuring the sliding plane is computed correctly, and subtracting two points in the sliding plane (*newDestinationPoint* and the modified *intersectionPoint*) succeeds at yielding a new velocity vector parallel to the plane. Nevertheless, a more straightforward solution exists. We can simplify and improve the numerical robustness of the algorithm by use of two points called the *near point* and *touch point*. Let the *touch point* be the position of the sphere's center at the time it would first touch the triangle if allowed to travel freely. Let the *near point* be the position of the sphere's center a moment before it first touches the triangle. (This is where we will actually position the sphere, analogous to the *newBasePoint* in Table I.) This is how our revised algorithm computes the *touch point* and *near point,* where $t \in [0, 1]$ is the time of the collision, calculated by the *collision detection step*:

$$touch\ point = source + \mathbf{vel} * t \quad (3)$$

$$distance = ||\mathbf{vel}|| * t \quad (4)$$

$$short\ distance = \max(distance - \varepsilon, 0) \quad (5)$$

$$near\ point = source + \widehat{\mathbf{vel}} * short\ distance \quad (6)$$

The shortened distance will move us arbitrarily close to the triangle, determined by a tolerance value $\varepsilon > 0$. Forcing this shortened distance to be at least zero prevents the sphere from moving backwards if *distance* is smaller than $\varepsilon$. With these points defined, we can provide a better way to find the sliding plane. Fauerby did something that made some sense conceptually, but is actually unnecessary: he used the *near point* (called *newBasePoint* in Table I) to compute the sliding plane's normal instead of the *touch point*, and consequently had to move the *intersection point* in the direction of –**vel** by *veryCloseDistance* so the points remain one unit from each other. When computed this way, the sliding plane will be tangent to the sphere when the sphere is at the *near point*, not when it is at the *touch point*. What is interesting though, is that this trick does not affect the sliding plane's normal at all! The sphere and sliding plane are moved backwards in concert with each other, so the distance between the sphere's center and the modified *intersection point* remains the same. The sliding plane's normal is what we really care about (it allows us to find the direction to slide in) and the precision gained in the plane's position is irrelevant. Therefore, we can simply use the unmodified intersection point and the vector pointing from the unmodified intersection point to the *touch point* to find the sliding plane. This makes the code simpler, easier to understand, faster, and more numerically robust. Specifically, (3-6) allow us to eliminate the `if` statement from line 14 of the algorithm from Table I and streamline our new algorithm.

### B. The Jittery Problem

The algorithm from Table I checks for collision against all triangles every time a new velocity is computed to ensure there are no obstacles. Unfortunately, if more than one triangle is hit per frame, the original algorithm does not correctly compute the final destination. Consider for instance the scenario depicted in Fig. 2, where a sphere approaches an obtuse terrain corner from the top coming down.

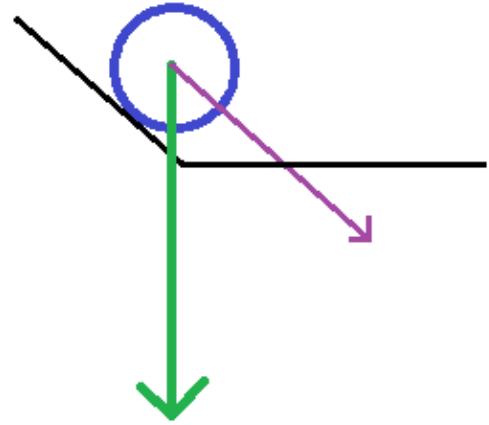

Fig. 2 Sphere approaching an obtuse corner. Incident to the center of the sphere are depicted the velocity vector (vertical) and its projection (oblique).

In this situation, the projected velocity vector will be parallel to the oblique triangle, and lead the sphere straight into the corner, as expected. However, this first iteration establishes an invariant that the second iteration fails to respect: once the sphere hits the first triangle, its motion *should* be restricted to the sliding plane, losing a degree of freedom. The original algorithm, however, does not propagate this information over to the second iteration. Consequently, the new velocity vector leads the sphere into the second triangle; so, to prevent a collision, the original algorithm projects the velocity vector onto a *new* sliding plane. Projecting the velocity onto a second sliding plane leads the sphere away from the corner when Fig. 2 suggests it should stop at the corner, as illustrated in Fig. 3:

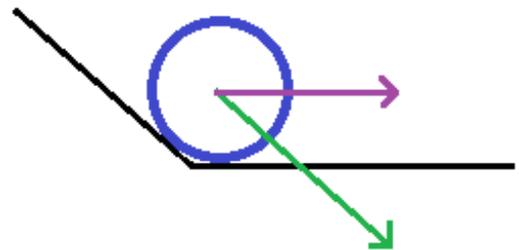

Fig. 3 Sphere moving away from the corner it was intended to stop at. Incident to the center of the sphere are depicted the velocity vector (oblique) and its projection (horizontal).

If the desired velocity vector is pointing towards the corner for several frames, the sphere will continue to try to go towards the corner, but its final destination will always be a small distance away from it. Consequently, the motion will appear very jittery; the sphere never settles down and becomes motionless in corners, creases, or dead ends.

### C. The Freezing Problem

Acute corners can cause even worse problems. Unless we arbitrarily cap the number of iterations (as Fauerby does at line 3 of Table I), the algorithm may run hundreds of times and adversely affect performance.

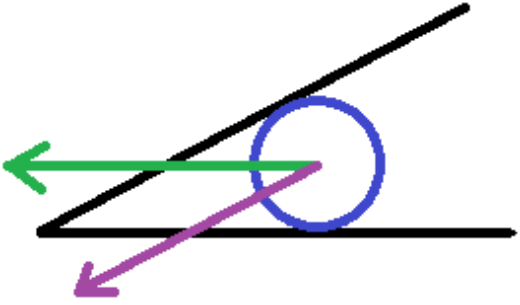

Fig. 4 Sphere stuck in an acute corner. Incident to the center of the sphere are depicted the velocity vector (horizontal) and its projection (oblique).

Imagine Fig. 4 with a very small angle between the two triangles. We find that the sphere intersects the oblique triangle almost immediately (i.e., at a *t* value very close to zero). When this occurs, the projected velocity vector will be almost the same length as the original velocity vector, and will lead the sphere into the horizontal triangle. Again, we find that the sphere will intersect this second triangle almost immediately. Consequently, the algorithm will continue to alternate between calculating collisions between the horizontal and oblique triangles until the velocity vector reaches zero length and we can finally "move" to our destination without encountering any obstacles. Without a cap on the number of iterations, this will cause the program to freeze or cause a stack overflow. We will soon see an improvement to this algorithm that guarantees it will stop after three iterations. The algorithm presented in Table I has a worst-case performance of five iterations [2], so our improved version yields better performance.

This freezing problem does not exist in isolation. Walking into corners like this (forcing multiple iterations to occur) combines particularly badly with other numerical robustness bugs, causing the sphere to penetrate the mesh. We suspect that floating point round-off errors become amplified with each iteration the algorithm is stuck in such a state, but these problems occur even when the number of iterations is capped.

### D. Reducing the Degrees of Freedom

Next, we look at methods to solve the problems unearthed in the two previous sections. Let us consider what happens when we project the velocity vector. When the sphere hits a triangle, we declare that the motion will (until the next frame) be constrained to the sliding plane. But, the algorithm from Table I is only aware of one sliding plane at a time. To prevent the sphere from leaving any of these sliding planes, we must remember all triangles the sphere collided with during this frame, and project the velocity vector onto the sliding planes of them all. With knowledge of the sliding planes from previous iterations, it is possible to correctly constrain the motion of the sphere using the correct number of degrees of freedom and eliminate jittery, unstable collision detection. Once we move the sphere to the *near point*, a simple case analysis on the number of previous iterations determines how to calculate a new velocity and destination.

### E. Colliding with One Plane

In the case of only one sliding plane, the calculations for the new velocity vector are similar to the ones in the original algorithm (Table I). They reduce to first projecting the sphere's destination onto the sliding plane, then moving it just over one unit away from the sliding plane so the sphere lies on the same side of the plane it came from. We use a *long radius* of $1+\varepsilon$, for some tolerance value $\varepsilon > 0$, to prevent the sphere from actually touching the sliding plane when it gets to its new destination. Recall that we applied similar logic when finding the *near point*.

$$long\ radius = 1 + \varepsilon \quad (7)$$

$$new\ dest = dest - (\text{plane\_dist}(sliding\ plane, dest) - long\ radius) \cdot \textbf{plane normal} \quad (8)$$

$$\textbf{new vel} = new\ dest - near\ point \quad (9)$$

Fauerby's original algorithm did not need to add a tolerance here because he placed the sliding plane a small distance away from the triangle.

### F. Colliding with Two Planes

In the case of two (non-parallel) sliding planes, we must constrain the motion of the sphere to the crease created by the intersection of the two planes. A crease vector of two planes is parallel to both, so we can start by finding the cross product of the two planes' normals. This produces a vector perpendicular to both planes' normals and consequently runs parallel to them. We then normalize this vector to get the crease vector, and project the remainder of the old velocity onto the crease vector to yield a signed distance. The signed distance tells us how far along the crease vector the new velocity is.

$$\textbf{p} = \textbf{first plane normal} \times \textbf{second plane normal} \quad (10)$$

$$\textbf{crease} = \hat{\textbf{p}} \quad (11)$$

$$signed\ distance = (dest - near\ point) \cdot \textbf{crease} \quad (12)$$

$$\textbf{new vel} = signed\ distance * \textbf{crease} \quad (13)$$

$$\text{new dest} = \text{near point} + \textbf{new vel} \qquad (14)$$

The sphere does not need to be explicitly kept a minimum distance away from the planes in this case. The sphere will maintain its distance from both planes provided it stops at the *near point* and then moves parallel to them.

### G. Colliding with Three Planes

In the case of three such sliding planes, (no two of which are parallel to each other), we have no degrees of freedom left. The intersection of three such planes is a point, so the sphere's motion must cease. The projected velocity becomes zero and we can skip calculating the sliding plane altogether:

$$\textbf{new vel} = (0,0,0) \qquad (15)$$

$$\text{new dest} = \text{near point} \qquad (16)$$

By keeping track of previously calculated sliding planes, we obtain smooth motion along creases and motionlessness in corners, putting an end to jittering. Recall that we have also imposed a cap on the number of iterations, meaning no type of geometry can cause the new algorithm to freeze or degrade performance. This time though, the cap is not arbitrary: we derive the bound of three from the fact that the sphere has three degrees of freedom. The original algorithm also contained an extra end condition at line 26: if the velocity vector was very small, the collision response algorithm would end. Under the new constraint that the algorithm will always terminate after at most three iterations, we can safely omit this check and further streamline our code.

### H. Two Parameters Are Not Enough

Our implementation (section I) reveals one important source of floating point round-off error present in the original algorithm (Table I) – the type that causes the sphere to intersect the mesh even though the equations appear correct. When written recursively, our improved algorithm has three parameters for the sphere's movement: the source position, the velocity, and the destination. This may seem surprising and redundant at first: if we know the source and velocity, we can derive the destination. Nevertheless, information may be lost: if the sum of two floating point numbers *a* and *b* is *c*, there is no guarantee that *c – b* is equal to *a,* because rounding may occur when calculating c. In the case of one sliding plane, we find the new velocity by subtracting the near point (i.e., the new source position) from the new destination. If we were to pass only the source and velocity as arguments to the next recursion, the function would have to re-derive the destination from these values, an opportunity for round-off error. There are two ways to solve this problem. The first is to pass all three values as arguments to make sure no unnecessary rounding occurs between calls. The second is to write the algorithm iteratively, making explicit passing of these values unnecessary. We cannot simply rewrite (7-9) to produce the new velocity directly and skip computing the new destination.

Doing so would complicate the equations and introduce more opportunities for round-off error.

### I. Implementation

This is a sample implementation of the revised algorithm. Unlike the tail recursive implementation from Table I, here we show an iterative implementation, which has the minor advantage of not needing to declare a "new" and "old" velocity and destination explicitly.

TABLE II
Our Improved Collision Response Step

```
VECTOR sphere_sweep(VECTOR pos, VECTOR vel)
1.  VECTOR dest = pos + vel;
2.  PLANE first_plane;

3.  for (int i = 0; i < 3; ++i) {
4.      CollisionStats stats = world->checkCollision(pos, vel);
5.      if (!stats.hit) return dest; // no collision
6.
7.      float dist = vel.length() * stats.t;
8.      float short_dist = max(dist - very_close_dist, 0.0f);
9.      pos += vel.normal() * short_dist;

10.     if (i == 0) {
11.         float long_radius = 1.0f + very_close_dist;
12.         first_plane = stats.sliding_plane();
13.         dest -= (plane_dist(first_plane, dest) - long_radius) * first_plane.n;
14.         vel = dest - pos;
15.     } else if (i == 1) {
16.         PLANE second_plane = stats.sliding_plane();
17.         VECTOR crease = first_plane.n.cross(second_plane.n).normal();
18.         float dis = (dest - pos).dot(crease);
19.         vel = dis * crease;
20.         dest = pos + vel;
        }
    }
21. return pos;
```

## IV. OPEN PROBLEMS AND CONCLUSIONS

Collision detection is notoriously difficult. One of the difficulties (discussed in this paper) is unstable code because of round-off errors. In this paper, we offer some insight into the problems that contribute to an unstable collision detection implementation, along with a few solution ideas.

We implemented the improved algorithm from Table II, and were not able to observe a single instance of the sphere jittering or penetrating the triangle mesh. Whereas a careful implementation of this improved algorithm is very effective at maintaining the invariant that the sphere must never intersect the geometry, there are a few unsolved problems. We noticed that the sphere snags and often gets caught on the edges of triangles, which can be described as the opposite of the jitteriness problem. Tweaking the tolerance value $\varepsilon$ can mitigate this problem; however, a deeper understanding of the cause seems necessary here.

ACKNOWLEGMENT

Jeff Linahan thanks Dr. Mirela Damian for proofreading and editing.